\documentclass[epj]{webofc}
\usepackage[varg]{txfonts}   % Web of Conferences font
\usepackage{bm}
\newcommand{\bra}[1]{\left\langle #1\right|}
\newcommand{\ket}[1]{\left| #1\right\rangle}
\usepackage{amsfonts}
\woctitle{CNR*15}
\begin{document}
\title{Forging the link between nuclear reactions and nuclear structure}

\author{W. H. Dickhoff\inst{1}\fnsep\thanks{\email{wimd@wuphys.wustl.edu}} %\and
       % Second author\inst{2}\fnsep\thanks{\email{Mail address for second
    %         author if necessary}} \and
    %    Third author\inst{3}\fnsep\thanks{\email{Mail address for last
    %         author if necessary}}
        % etc.
}

\institute{Department of Physics, Washington University, St. Louis, Missouri 63130, USA 
%\and
   %        the second here 
%\and
   %        Last address
          }

\abstract{%
A review of the recent applications of the dispersive optical model (DOM) is presented.
Emphasis is on the nonlocal implementation of the DOM that is capable of describing ground-state properties accurately when data like the nuclear charge density are available.
The present understanding of the role of short- and long-range physics in 
determining proton properties near the Fermi energy for stable closed-shell 
nuclei has relied mostly on data from the $(e,e'p)$ reaction. 
Hadronic tools to extract such spectroscopic information have been hampered
by the lack of a consistent reaction description that provides unambiguous and
\emph{undisputed} results.
The DOM, conceived by Claude Mahaux, 
provides a unified description of both elastic nucleon scattering and 
structure information related to single-particle properties below the Fermi 
energy.
We have recently introduced a nonlocal dispersive optical potential for both the real and imaginary part.
Nonlocal absorptive potentials yield equivalent elastic differential cross sections for ${}^{40}$Ca as compared to local ones but change the $\ell$-dependent absorption profile suggesting important consequences for the analysis of nuclear reactions.
Below the Fermi energy, nonlocality is essential for an accurate representation of particle number and the nuclear charge density. Spectral properties implied by $(e,e'p)$ and $(p,2p)$ reactions are correctly described, including the energy distribution of about 10\% high-momentum protons obtained at Jefferson Lab.
The nonlocal DOM allows a complete description of experimental data both above (up to 200 MeV) and below the Fermi energy in $^{40}$Ca. It is further demonstrated that elastic nucleon-nucleus scattering data constrain the spectral strength in the continuum of orbits that are nominally bound in the independent-particle model.
Extension of this analysis to $^{48}$Ca allows a prediction of the neutron skin of this nucleus that is larger than most predictions made so far.
}
\maketitle
\section{Introduction}
\label{intro}
The properties of a nucleon that is strongly influenced by the presence of other nucleons have traditionally been studied in separate energy domains and frameworks.
Positive energy nucleons are described by fitted optical potentials mostly in local form~\cite{Varner91,Koning03} which generate the distorted waves employed in the analysis of nuclear reactions.
Bound nucleons have been analyzed with static potentials.
At first, this lead to an independent-particle model (IPM) which is then modified by the interaction between valence nucleons in a very limited configuration space as in traditional shell-model calculations~\cite{Brown01,Caurier05}.
No attention is paid in this approach to the larger energy domain of giant resonances or effects associated with fast nucleons.
These high-momentum nucleons have been experimentally identified as playing a role in the properties of nuclei and play a very important role in the success of \textit{ab initio} methods to describe nuclei that rely on interactions with substantial repulsive cores. 
The link between nuclear reactions and nuclear structure is however possible by considering these potentials as representing  different energy domains of one underlying nucleon self-energy provided by the framework of Green's function theory~\cite{Dickhoff08}.
This idea was implemented in the DOM by Mahaux and Sartor~\cite{Mahaux91}.
By employing dispersion relations, the method provides a critical link between the physics above and below the Fermi energy with both sides being influenced by the absorptive potentials on the other side.
Such a link is not present in traditional optical potentials that have been constructed by independently fitting the real and imaginary parts.

When a Hamiltonian for the description of nonrelativistic nucleons is employed as a starting point for the description of nuclei, a perturbation expansion can be derived for the nucleon single-particle propagator~\cite{Dickhoff08}.
A compact formulation for the propagator leads to the Dyson equation which relates the propagator to a noninteracting one used as a starting point and the so-called self-energy.
The latter acts as the potential in a Schr{\"{o}}dinger-like equation derived from the Dyson equation with solutions that identify the properties of nucleons that are either removed or added to a ground-state nucleus.
In the latter case, the solutions describe all aspects of elastic proton or neutron scattering from a target ground state.
Removal properties can be employed to generate expectation values of one-body operators in the ground state like densities, as well as a contribution to the ground-state energy when it is dominated by the effects of two-body interactions.
The potential (self-energy) is energy dependent, complex in certain energy domains, and its real and imaginary part obey a dispersion relation.
Furthermore, at the Hartree-Fock (HF) level the self-energy is already nonlocal and all theoretical calculations of the self-energy confirm that both the real and imaginary part of the self-energy are nonlocal.
The traditional use of the optical potential as local and nondispersive is therefore a gross misrepresentation of the actual physics and is in need of updating to a nonlocal and dispersive implementation.
The work discussed in this contribution represents a step in this direction.

While the DOM provides an intermediary between theory and experiment, the ultimate goal is to be able to accurately calculate the nucleon self-energy in an \textit{ab initio} manner. Currently we are very far from this goal but it should be in our sights with the help of the experience gained from the DOM.
In Sec.~\ref{sec-1} a brief outline of the ingredients of the DOM are presented. 
Results are presented for a fully nonlocal implementation of the DOM for ${}^{40}$Ca in Sec.~\ref{sec-2} including results for the particle spectral function that quantitatively clarify the removal of single-particle (sp) strength from mean-field orbits.
In Sec.~\ref{sec-3} recent results for ${}^{48}$Ca are presented with emphasis on the neutron skin which emerges as a quantity that can be extracted from the fit to the available data when results for ${}^{40}$Ca are employed as a starting point.
Some brief conclusions and outlook are provided in Sec.~\ref{sec-4}. 

\section{Green's functions and the dispersive optical model}
\label{sec-1}
The self-energy $\Sigma_{\ell j}$ provides the critical ingredient to solve the Dyson equation for the nucleon propagator $G_{\ell j}$.  
Employing an angular momentum basis, it reads
\begin{eqnarray}
\label{eq:dyson}
G_{\ell j}(k,k';E) = \frac{\delta(k-k')}{k^2}G^{(0)}(k;E) + G^{(0)}(k;E) \int \!\! dq\ q^2\ 
\Sigma_{\ell j}(k,q;E)\ G_{\ell j}(q,k';E) .
%\nonumber
\end{eqnarray}
The noninteracting propagators $G^{(0)}$ only contain kinetic energy contributions.
The solution of this equation generates $S_{\ell j}(k;E) =   \textrm{Im}\ G_{\ell j}(k,k;E)/\pi$, the hole spectral density, 
%\begin{equation}
%\label{eq:holes}
%\end{equation}
for negative continuum energies.
The spectral strength at $E$, for a given $\ell j$, is given by 
\begin{equation}
S_{\ell j}(E) = \int_{0}^\infty dk\ k^2\ S_{\ell j}(k;E) .
\label{eq:specs}
\end{equation}
For discrete energies one solves the eigenvalue equation for the overlap functions
\begin{equation}
\phi^n_{\ell j}(k) = \bra{\Psi^{A-1}_n}a_{k \ell j} \ket{\Psi^A_0},
\label{eq:overlap}
\end{equation}
%\begin{equation}
%\psi^n_{\ell j}(r) = \bra{\Psi^{A-1}_n}a_{r \ell j} \ket{\Psi^A_0} ,
%\label{eq:overlap}
%\end{equation}
for the removal of a nucleon with momentum $k$ and discrete quantum numbers $\ell$ and $j$~\cite{Dickhoff10}.
The removal energy corresponds to
%\begin{equation}
$\varepsilon^-_n=E^A_0 -E^{A-1}_n$ 
%\label{eq:eig}
%\end{equation}
with the normalization for such a solution $\alpha_{qh}$ given by
\begin{equation}
S^n_{\ell j} = \left( 1 - \left. \frac{
\partial \Sigma_{\ell j}(\alpha_{qh},
\alpha_{qh}; E)}{\partial E} \right|_{\varepsilon^-_n}\right)^{-1},
\label{eq:sfac}
\end{equation}
which is referred to as the spectroscopic factor.
We note that from the solution of the Dyson equation below the Fermi energy, one can generate the one-body density matrix by integrating the non-diagonal imaginary part of the propagator up to the Fermi energy and therefore access the expectation values of one-body operators in the ground state including particle number, kinetic energy and charge density~\cite{Dickhoff08}. 
The latter is obtained by folding the point density with the nucleon form factors~\cite{Brown79}.
For positive energies, it was already realized long ago that the reducible self-energy provides the scattering amplitude for elastic nucleon scattering~\cite{Bell59}.

The self-energy fulfills the dispersion relation which relates the physics of bound nucleons to those that propagate at positive energy~\cite{Dickhoff08}.
It contains a
%\begin{eqnarray} 
%\mbox{Re}\ \Sigma_{\ell j}(r,r';E)\!& =& \! \Sigma^s_{\ell j} (r,r')\! - \! {\cal P} \!\!
%\int_{\varepsilon_T^+}^{\infty} \!\! \frac{dE'}{\pi} \frac{\mbox{Im}\ \Sigma_{\ell j}(r,r';E')}{E-E'}  \nonumber \\
%&+&{\cal P} \!\!
%\int_{-\infty}^{\varepsilon_T^-} \!\! \frac{dE'}{\pi} \frac{\mbox{Im}\ \Sigma_{\ell j}(r,r';E')}{E-E'} ,
%\label{eq:disprel}
%\end{eqnarray}
static correlated HF term and dynamic parts representing the coupling in the $A\pm1$ systems that start and end at the Fermi energies for addition ($\varepsilon_F^+ = E^{A+1}_0-E^A_0$) and removal ($\varepsilon_F^-=E^A_0-E^{A-1}_0$), respectively.
The latter feature is particular to a finite system and allows for discrete quasi particle and hole solutions of the Dyson equation where the imaginary part of the self-energy vanishes.
It is convenient to introduce the average Fermi energy
%\begin{equation}
$\varepsilon_F = \frac{1}{2} \left[
\varepsilon_F^+  - \varepsilon_F^- \right]$
%\label{eq:FE}
%\end{equation}
and employ the subtracted form of the dispersion relation calculated at this energy~\cite{Mahaux91,Dickhoff10}
\begin{eqnarray} 
\mbox{Re}\ \Sigma_{\ell j}(k,k';E)\! = \!  \Sigma_{\ell j} (k,k';\varepsilon_F) % \hspace{3.0cm}
- \! {\cal P} \!\!
\int_{\varepsilon_F^+}^{\infty} \!\! \frac{dE'}{\pi} \mbox{Im}\ \Sigma_{\ell j}(k,k';E') \left[ \frac{1}{E-E'}  - \frac{1}{\varepsilon_F -E'} \right]  \nonumber  \\
+{\cal P} \!\!
\int_{-\infty}^{\varepsilon_F^-} \!\! \frac{dE'}{\pi} \mbox{Im}\ \Sigma_{\ell j}(k,k';E') \left[ \frac{1}{E-E'}
-\frac{1}{\varepsilon_F -E'} \right]  ,
 \label{eq:sdisprel} 
\end{eqnarray}
where $\mathcal{P}$ represents the principal value.
The beauty of this representation was recognized by Mahaux and Sartor~\cite{Mahaux86,Mahaux91} since it provides a link with empirical information both for the real part of the nonlocal self-energy at the Fermi energy (probed by a multitude of HF calculations) as well as through empirical knowledge of the imaginary part of the optical potential also constrained by experimental data.
Consequently Eq.~(\ref{eq:sdisprel}) yields a dynamic contribution to the real part linking \textrm{both} energy domains around the Fermi energy.
Empirical information near $\varepsilon_F$ is emphasized by Eq.~(\ref{eq:sdisprel}) because of the $E'^{-2}$-weighting in the integrands.
The real self-energy at the Fermi energy will be denoted by $\Sigma_{HF}$.

Early implementations of the DOM employed a local form of the imaginary part of the potential and transformed the nonlocality of $\Sigma_{HF}$ to a local energy-dependent potential~\cite{Mahaux91}.
Traditional volume and surface absorption terms with Woods-Saxon form factors were employed.
Initial revival of the DOM in St. Louis contained the idea that a simultaneous fit to ${}^{40}$Ca and ${}^{48}$Ca allows an extrapolation to more neutron-rich calcium isotopes generating a data-driven extrapolation to the drip line~\cite{Charity06,Charity07}.
As will be seen later, a clear motivation to generate more neutron scattering data on ${}^{48}$Ca emerged from this work culminating in a comprehensive DOM article for several relevant parts of the chart of nuclides~\cite{Charity11} while still employing local potentials.

The DOM thus provides an ideal strategy to predict properties for exotic nuclei by utilizing extrapolations of the potentials towards the respective drip lines~\cite{Charity06,Charity11,Charity14}.
While it is possible to correct the artificial energy dependence of the HF self-energy for some properties near the Fermi energy like spectroscopic factors,
it was necessary to employ the approximate expressions of limited accuracy for the properties of nucleons below the Fermi energy like occupation numbers that were developed by Mahaux and Sartor~\cite{Mahaux91} .
By restoring the proper treatment of nonlocality in the HF contribution, it was possible to overcome this problem~\cite{Dickhoff10} although the local treatment of the absorptive potentials yielded a poor description of the nuclear charge density and particle number.

Recently a fully nonlocal treatment of these potentials for the nucleus ${}^{40}$Ca was presented~\cite{Mahzoon14} with the aim to include \textit{all} available data below the Fermi energy that can be linked to the nucleon single-particle propagator~\cite{Dickhoff08} while maintaining a correct description of the elastic-scattering data. 
The result is a DOM potential that can be interpreted as the nucleon self-energy constrained by all available experimental data up to 200 MeV.
Such a self-energy allows for a consistent treatment of nuclear reactions that depend on distorted waves generated by optical potentials as well as overlap functions and their normalization for the addition and removal of nucleons to discrete final states. 
The re-analysis of such reactions may further improve the consistency of the extracted structure information.
Extending this version of the DOM to $N \ne Z$ allows for predictions of properties that require the simultaneous knowledge of both reaction and structure information since at present few weakly-interacting probes are available for exotic nuclei~\cite{Dickhoff10a}. 

\section{Results for ${}^{40}$Ca}
\label{sec-2}
The nonlocal DOM potential generated in Ref.~\cite{Mahzoon14} was motivated by theoretical calculations of the self-energy for ${}^{40}$Ca which either emphasized long-range correlations (LRC)~\cite{seth11} or short-range correlations (SRC)~\cite{Dussan11}.
In Ref.~\cite{seth11} the results of the \textit{ab initio} Faddeev random phase approximation (FRPA) which emphasizes the coupling of nucleons to low-lying collective states and to giant resonances, were compared with local DOM results.
The nonlocality of the DOM potentials was represented by simple Gaussians which allow an analytic projection onto orbital angular momentum.
We can now compare volume integrals for the potentials generated in Ref.~\cite{Mahzoon14} with these \textit{ab initio} results.
\begin{figure}
\centering
\includegraphics[width=12cm,clip]{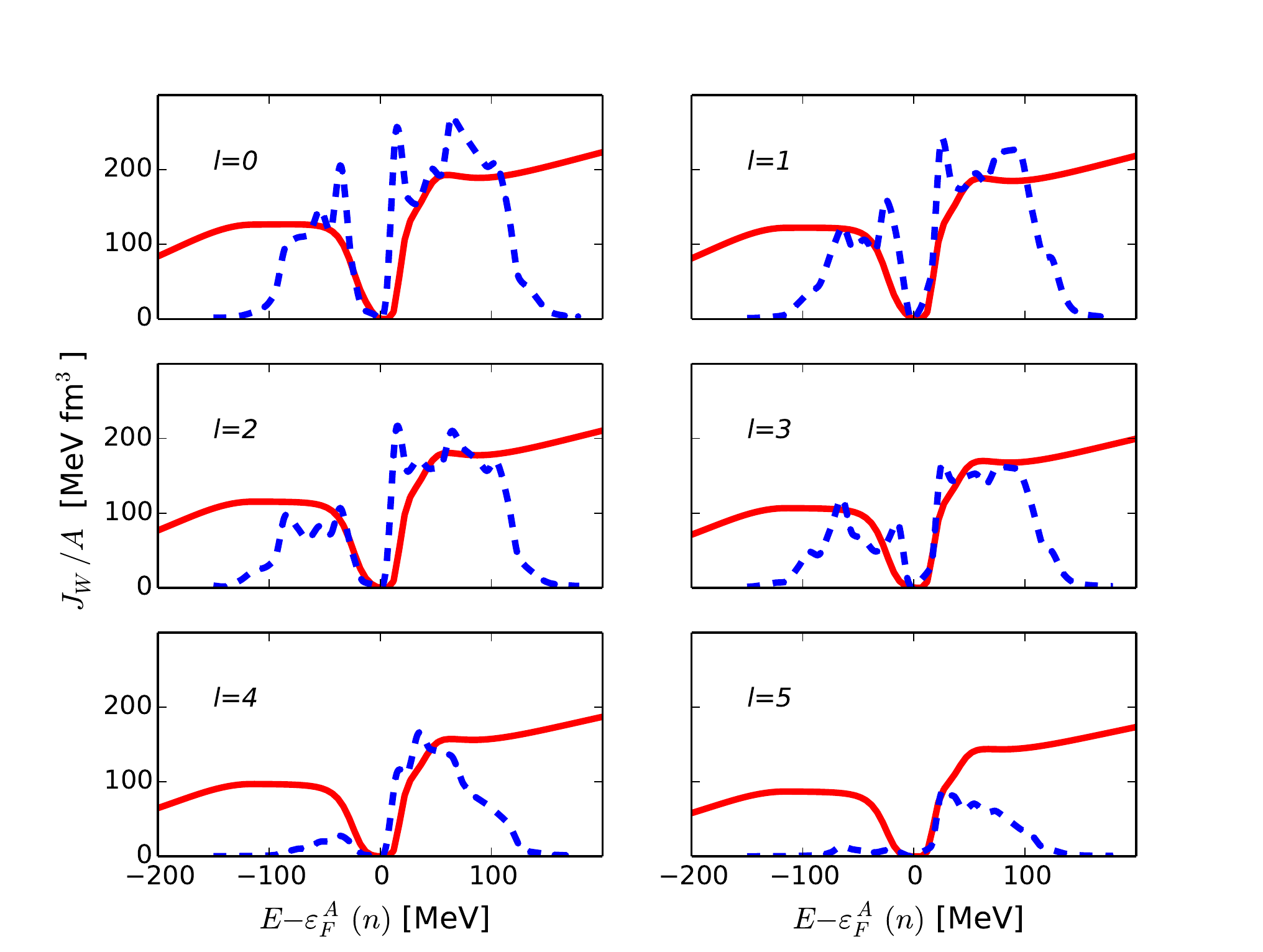}
\caption{Comparison of volume integrals for different orbital angular momenta for the nonlocal DOM (solid)  and FRPA (dashed) calculations in a wide energy domain. The curves are centered at the neutron Fermi energy for ${}^{40}$Ca.}
\label{fig-1}       % Give a unique label
\end{figure}
In Fig.~\ref{fig-1} we compare these volume integrals for the imaginary part of the potentials as a function of energy in a domain which spans 200 MeV on either side of the Fermi energy for several orbital angular momenta~\cite{Hossein15}.
We observe that the DOM results have been constrained by level information, the experimental charge density,high-momentum removal data from Jefferson Lab~\cite{Rohe04,Rohe04A} extending to very negative energies as well as available elastic scattering data up to 200 MeV.
The FRPA calculations were performed in a limited configuration space which is clearly illustrated in Fig.~\ref{fig-1}.
We note that the agreement between DOM and FRPA results is particularly good in the vicinity of the Fermi energy but the quality of the FRPA calculations is limited to about 60 MeV on either side of the Fermi and also deteriorates with increasing angular momentum illustrating the limitations of the configuration space.

\begin{figure}
\centering
\includegraphics[width=9cm,clip]{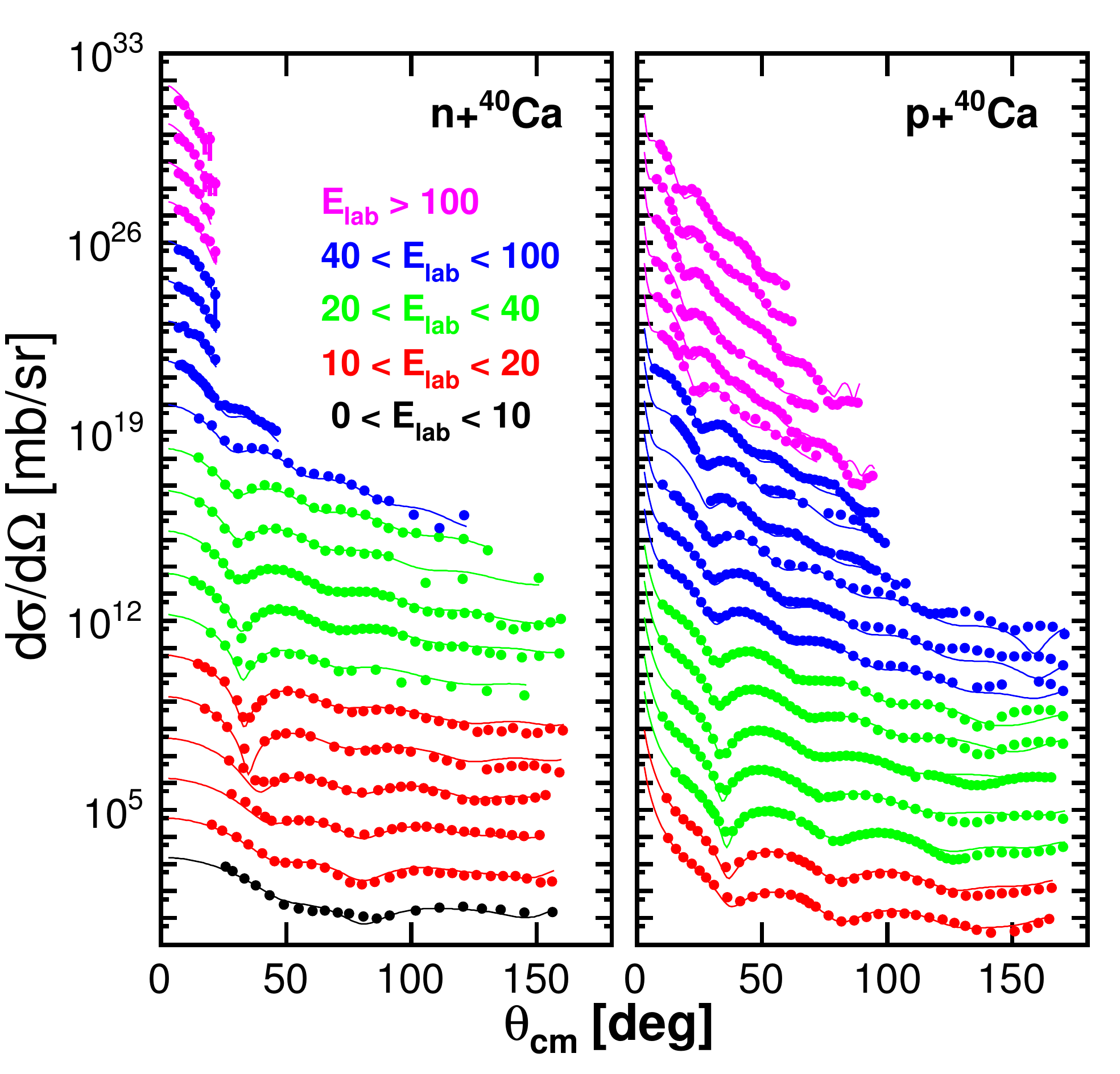}
\caption{Calculated and experimental elastic-scattering angular distributions of the differential cross section for both$n+^{40}$Ca and $p+^{40}$Ca. Data for each energy are offset for clarity. References to the data can be found in Ref.~\cite{Charity11}.}
\label{fig-2}       % Give a unique label
\end{figure}
\begin{figure}
\centering
\includegraphics[width=7cm,clip]{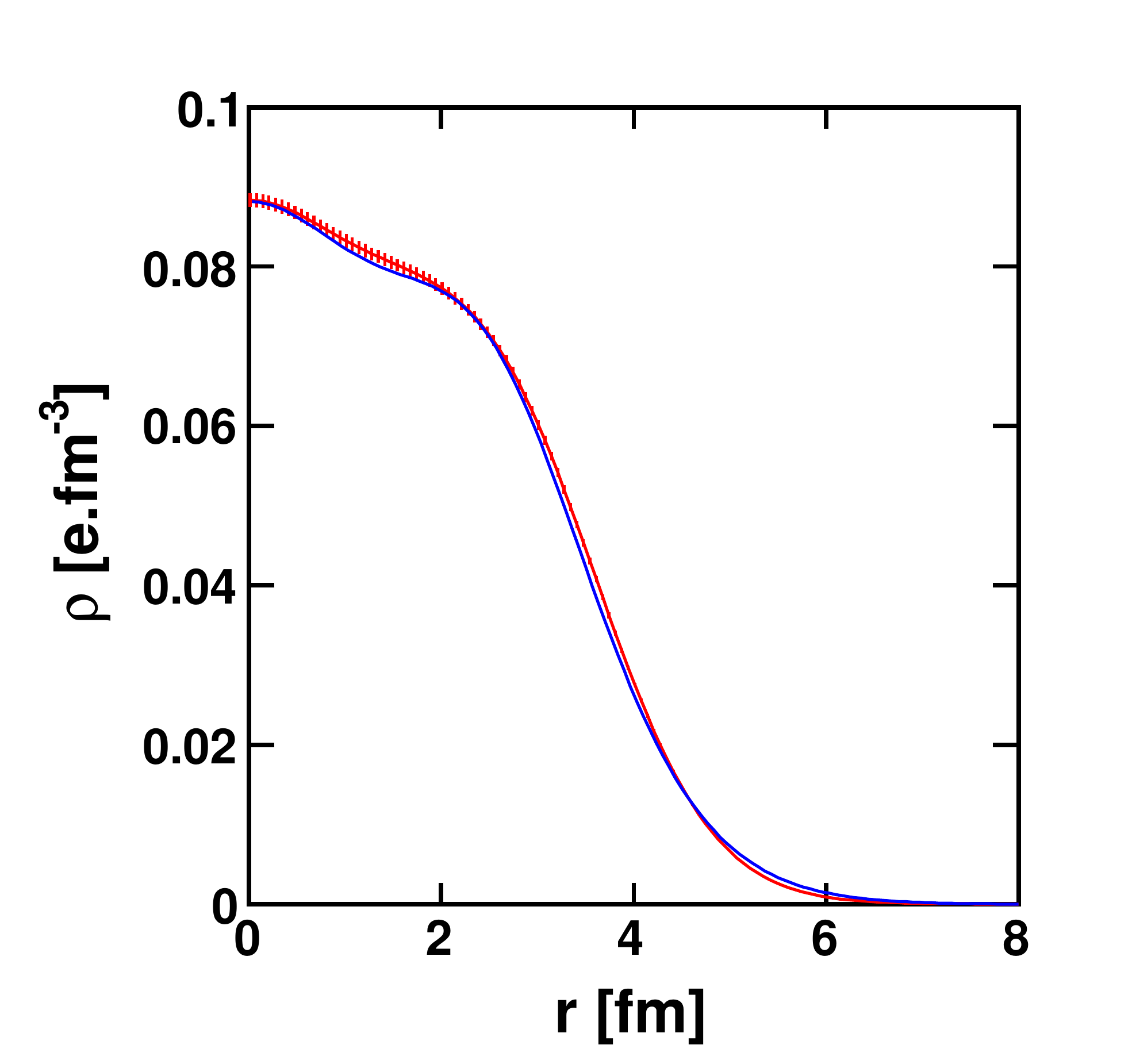}
\caption{Comparison of experimental charge density of ${}^{40}$Ca~\cite{deVries1987} (thick hashed line representing a 1\% error) with the DOM fit (solid curve).}
\label{fig-3}       % Give a unique label
\end{figure}

The traditional good quality of DOM fits of elastic scattering data is maintained with the nonlocal implementation as illustrated in Fig.~\ref{fig-2} for both proton and neutron elastic scattering~\cite{Mahzoon14}.
Similar statements pertain to the reaction cross section for protons and total and reaction cross section for neutrons~\cite{Mahzoon14}.
The quality of the description of data below the Fermi energy is however dramatically improved over earlier DOM calculations with a local absorptive potential~\cite{Dickhoff10}.
In Fig.~\ref{fig-3} the experimental charge density of ${}^{40}$Ca~\cite{deVries1987} is compared with the DOM fit~\cite{Mahzoon14}.
The accurate description of the charge density can only be obtained by incorporating a nonlocal absorptive potential which decreases the admixture of higher orbital angular momentum components in the ground state sufficiently fast to allow particle number to be described properly.
We note that local absorptive potentials exhibit no such orbital angular momentum dependence and overestimate particle number substantially~\cite{Dickhoff10}.

Since elastic scattering data only depend on the asymptotic form of the elastic scattering wave function, it is important to realize that by employing the nonlocal DOM in a wide energy domain, it is possible to assess where the spectral strength of mostly occupied states resides in the continuum~\cite{Dussan14}.
The corresponding particle spectral function for a finite system requires the complete reducible self-energy $\Sigma^{\text{red}}$.
This reducible self-energy is calculated by employing a momentum-space scattering code discussed in Ref.~\cite{Dussan11}.
In an angular-momentum basis, iterating the irreducible self-energy $\Sigma$ to all orders, yields
\begin{eqnarray}\label{eq:redSigma1}
\Sigma_{\ell j}^{\text{red}}(k,k^\prime ;E)  =  \Sigma_{\ell j}(k,k^\prime ;E)
  +  \!\!       \int \!\! dq q^2\ \Sigma_{\ell j}(k,q;E)\ G^{(0)}(q;E )\ \Sigma_{\ell j}^{\text{red}}(q,k^\prime ;E) ,
\end{eqnarray}
where $G^{(0)}(q; E ) = (E - \hbar^2q^2/2m + i\eta)^{-1}$ is again the free propagator.
The propagator is then obtained from the Dyson equation in the following form~\cite{Dickhoff08}
\begin{eqnarray}
G_{\ell j}(k, k^{\prime}; E) = \frac{\delta( k - k^{\prime})}{k^2}G^{(0)}(k; E)  
 \label{eq:gdys1}  
		+	      G^{(0)}(k; E)\ \Sigma^{\text{red}}_{\ell j}(k, k^{\prime}; E)\ G^{(0)}(k'; E) .	
\end{eqnarray}
The on-shell matrix elements of the reducible self-energy in Eq.~(\ref{eq:redSigma1}) are sufficient to describe all aspects of elastic scattering like differential, reaction, and total cross sections as well as polarization data~\cite{Dussan11}.
While only this element is strictly related to data, the DOM analysis in its nonlocal implementation correctly and simultaneously describes all relevant data at all energies (up to 200 MeV) both below and above the Fermi energy~\cite{Mahzoon14}.
We therefore assume that the DOM reducible self-energy is reasonably unique and may therefore provide insight into the depletion of the Fermi sea.
The spectral representation of 
%\begin{eqnarray}
%G_{\ell j}(k ,k' ; E) 
%&= & \sum_m \frac{\bra{\Psi^A_0} a_{k\ell j}
%\ket{\Psi^{A+1}_m} \bra{\Psi^{A+1}_m} a^\dagger_{k' \ell j} \ket{\Psi^A_0}
%}{ E - (E^{A+1}_m - E^A_0 ) +i\eta }   \nonumber \\
%& + & \sum_n \frac{\bra{\Psi^A_0} a^\dagger_{k' \ell j} \ket{\Psi^{A-1}_n}
%\bra{\Psi^{A-1}_n} a_{k \ell j} \ket{\Psi^A_0} }{
%E - (E^A_0 - E^{A-1}_n) -i\eta} . 
%\nonumber %\label{eq:prop}
%\end{eqnarray}
the particle part of the propagator, referring to the $A+1$ system, appropriate for a treatment of the continuum and possible open channels is given by~\cite{Mahaux91}
\begin{eqnarray}
G_{\ell j}^{p}(k ,k' ; E)  =  
\sum_n  \frac{ \phi^{n+}_{\ell j}(k) \left[\phi^{n+}_{\ell j}(k')\right]^*
}{ E - E^{*A+1}_n +i\eta }   \label{eq:propp} 
 +  
\sum_c \int_{T_c}^{\infty} dE'\  \frac{\chi^{cE'}_{\ell j}(k) \left[\chi^{cE'}_{\ell j}(k')\right]^* }{
E - E' +i\eta} . 
\end{eqnarray}
Overlap functions for bound $A+1$ states are given by 
\begin{equation}
\phi^{n+}_{\ell j}(k)=\bra{\Psi^A_0} a_{k\ell j}
\ket{\Psi^{A+1}_n},
\end{equation}
 whereas those in the continuum are given by 
\begin{equation}
  \chi^{cE}_{\ell j}(k)=\bra{\Psi^A_0} a_{k\ell j} \ket{\Psi^{A+1}_{cE}}
  \end{equation}
indicating the relevant channel by $c$ and the energy by $E$.
Excitation energies in the $A+1$ system are with respect to the $A$-body ground state $E^{*A+1}_n = E^{A+1}_n -E^A_0$.
Each channel $c$ has an appropriate threshold indicated by $T_c$ which is the experimental threshold with respect to the ground-state energy of the $A$-body system.
The overlap function for the elastic channel can be explicitly calculated by solving the Dyson equation while it is also possible to obtain the complete spectral density for $E>0$ 
\begin{eqnarray}
S_{\ell j}^{p}(k ,k' ; E) 
=
%  \sum_n  \phi^{n+}_{\ell j}(k) \left[ \phi^{n+}_{\ell j}(k') \right]^* \delta \left( E- E^{*A+1}_n \right)
%  \nonumber \\
%& + & 
\sum_c \chi^{cE}_{\ell j}(k) \left[ \chi^{cE}_{\ell j}(k') \right]^* .
\label{eq:specp}
\end{eqnarray}
The overlap functions for bound particle states and their normalization (spectroscopic factor) can be calculated in the same fashion as the corresponding quantities for hole states~\cite{Dussan11}.
The spectral density at continuum energies is obtained from crossing the branch cut on the real axis.
In practice, this requires solving the scattering problem twice at each energy so that one may employ
\begin{eqnarray}
\!\! S_{\ell j}^{p}(k ,k' ; E) 
= \frac{i}{2\pi} \left[ G_{\ell j}^{p}(k ,k' ; E^+) - G_{\ell j}^{p}(k ,k' ; E^-) \right]
\label{eq:specpp}
\end{eqnarray}
with $E^\pm =E\pm i\eta$, and only the elastic-channel contribution to Eq.~(\ref{eq:specp}) is explicitly known.
Equivalent expressions pertain to the hole part of the propagator $G_{\ell j}^{h}$~\cite{Mahaux91}.

\begin{figure}[tbp]
\centering
\includegraphics[width=9cm,clip]{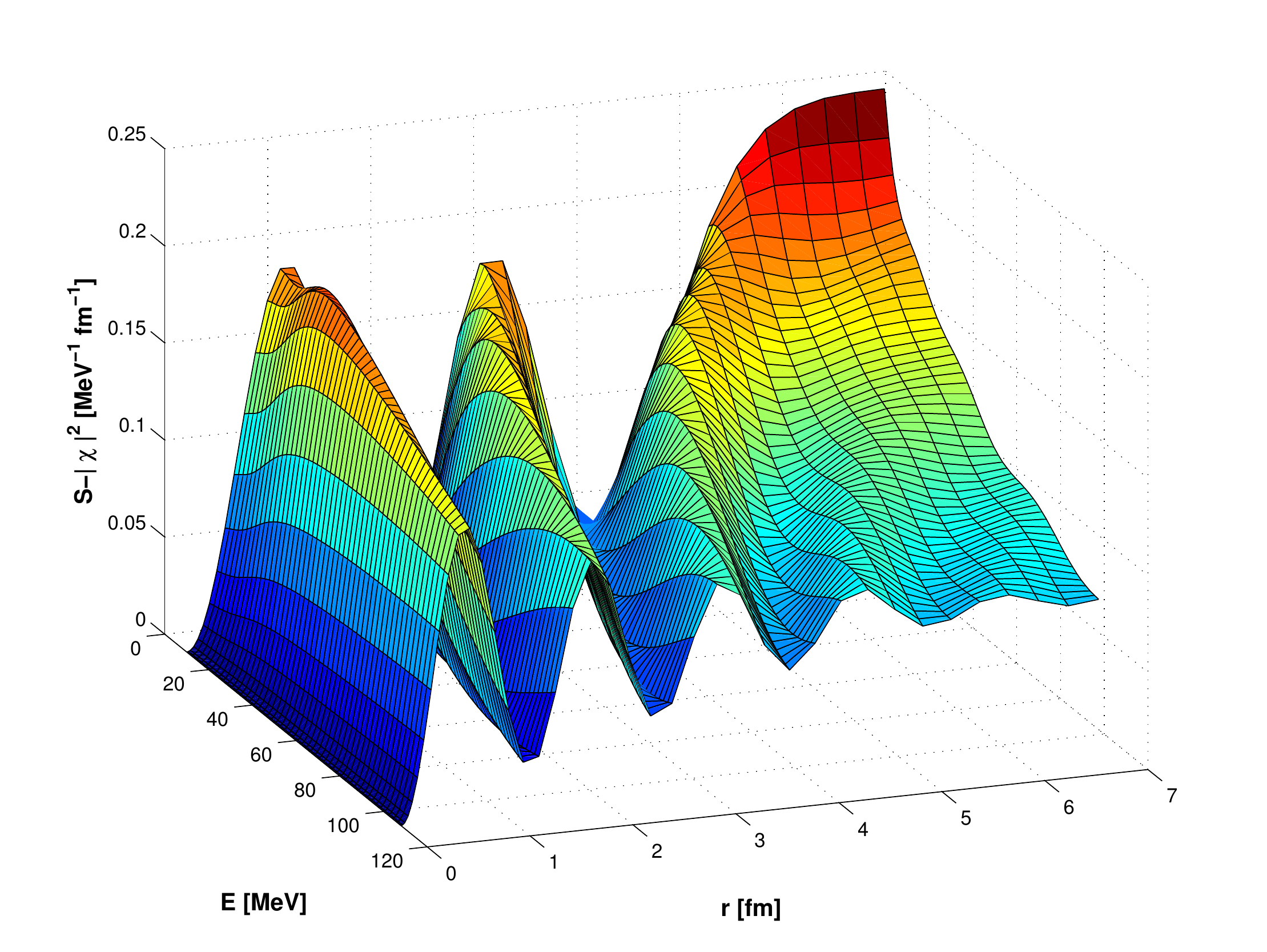}
\caption{Difference between the particle spectral function for $s_{1/2}$ and the contribution of the elastic-scattering wave function multiplied by $r^2$, as a function of both energy and position. Asymptotically with $r$, this difference is constant and determined only by the inelasticity.}
\label{fig:Sandpsi}
\end{figure}

The calculations are performed in momentum space according to Eq.~(\ref{eq:redSigma1}) to generate the off-shell reducible self-energy and thus the spectral density by employing Eqs.~(\ref{eq:gdys1}) and (\ref{eq:specpp}).
Because the momentum-space spectral density contains a delta-function associated with the free propagator, it is convenient for visualization purposes to consider a Fourier transform to coordinate space 
\begin{eqnarray}
S_{\ell j}^{p}(r ,r' ; E) = \frac{2}{\pi} \label{eq:specpr} 
  \int \!\! dk k^2 \! \int \!\! dk' k'^2 j_\ell(kr) S_{\ell j}^{p}(k ,k' ; E) j_\ell(k'r') ,
\end{eqnarray}
which has the physical interpretation for $r=r'$ as the probability density $S_{\ell j}(r;E)$ for adding a nucleon with energy $E$ at a distance $r$ from the origin for a given $\ell j$ combination.
By employing the asymptotic analysis to the propagator in coordinate space following \textit{e.g.} Ref.~\cite{Dickhoff08}, one may express the elastic-scattering wave function that contributes to Eq.~(\ref{eq:specp}) in terms of the half on-shell reducible self-energy obtained according to
\begin{eqnarray}
\chi^{el E}_{\ell j}(r)  = \left[ \frac{2mk_0}{\pi \hbar^2} \right]^{1/2} \left\{ j_\ell(k_0r)  \label{eq:elwf} 
 +   \int \!\! dk k^2 j_\ell(kr) G^{(0)}(k;E) \Sigma_{\ell j}(k,k_0;E) \right\} ,
\nonumber
\end{eqnarray}
where $k_0$ is related to the scattering energy in the usual way.
We subtract this contribution to the spectral function given by its absolute square from $S_{\ell j}(r;E)$ in Fig.~\ref{fig:Sandpsi} for different energies in the channel with $\ell =0, j=1/2$.
Asymptotically at large distances, the influence of other open channels is represented by an almost constant shift whereas, inside the range of the potential, a pattern related to the absorptive properties of the potential and the orbits that are occupied emerges.

The presence of strength in the continuum associated with mostly-occupied orbits (or mostly empty but $E<0$  orbits) is then obtained by double folding the spectral density in Eq.~(\ref{eq:specpr}) in the following way
\begin{eqnarray}
\!\!\! S_{\ell j}^{n+}(E) 
=  \int \!\! dr r^2 \!\! \int \!\! dr' r'^2 \phi^{n-}_{\ell j}(r) S_{\ell j}^{p}(r ,r' ; E) \phi^{n-}_{\ell j}(r') ,
\label{eq:specfunc}
\end{eqnarray}
using an overlap function 
\begin{equation}
\sqrt{S^n_{\ell j}} \phi^{n-}_{\ell j}(r)=\bra{\Psi^{A-1}_n} a_{r\ell j} \ket{\Psi^{A}_0} , 
\label{eq:overm}
\end{equation}
corresponding to a bound orbit with $S^n_{\ell j}$ the relevant spectroscopic factor and $\phi^{n-}_{\ell j}(r)$  normalized to 1~\cite{Dussan11}.
%We will refer to this quantity as the spectral strength of a bound orbit labeled by $n \ell j$. 
In the case of an orbit below the Fermi energy, this strength identifies where the depleted strength resides in the continuum.
The occupation number of this orbit is given by an integral over a corresponding folding of the hole spectral density
\begin{eqnarray}
\!\!\!\!\!\!\! S_{\ell j}^{n-}(E) 
= \!\! \int \!\! dr r^2 \!\! \int \!\! dr' r'^2 \phi^{n-}_{\ell j}(r) S_{\ell j}^{h}(r ,r' ; E) \phi^{n-}_{\ell j}(r') ,
\label{eq:spechr}
\end{eqnarray}
where $S_{\ell j}^{h}(r,r';E)$ provides equivalent information below the Fermi energy as $S_{\ell j}^{p}(r,r';E)$ above.
An important sum rule is valid for the sum of the occupation number for the orbit $n_{n \ell j}$ and its depletion number $d_{n \ell j}$~\cite{Dickhoff08}
\begin{eqnarray}
\!\! 1 =  n_{n \ell j} + d_{n \ell j} \!\!
\label{eq:sumr} 
 =\!\! \int_{-\infty}^{\varepsilon_F} \!\!\!\! dE\ S_{\ell j}^{n-}(E) \!+ \!\! \int_{\varepsilon_F}^{\infty} \!\!\!\! dE\ S_{\ell j}^{n+}(E)  ,%\nonumber
\end{eqnarray}
equivalent to $a^\dagger_{n \ell j} a_{n \ell j} +a_{n \ell j}a^\dagger_{n \ell j} =1$.
%The average Fermi energy $\varepsilon_F \equiv \frac{1}{2} \left[ (E^{A+1}_0-E^A_0) + (E^A_0 - E^{A-1}_0) \right]$  is introduced here~\cite{Mahaux91}.
Strength above $\varepsilon_F$, as expressed by Eq.~(\ref{eq:specfunc}), reflects the presence of the imaginary self-energy at positive energies.
Without it, the only contribution to the spectral function comes from the elastic channel.
The folding in Eq.~(\ref{eq:specfunc}) then involves integrals of orthogonal wave functions and yields zero.
%Equation~(\ref{eq:specfunc}) therefore identifies the presence of mostly occupied spectral strength in the continuum.
Because it is essential to describe elastic scattering with an imaginary potential, 
it automatically ensures that the elastic channel does not exhaust the spectral density and therefore some spectral strength associated with IPM bound orbits also occurs in the continuum.

\begin{figure}[tbp]
\centering
\includegraphics*[scale=0.4]{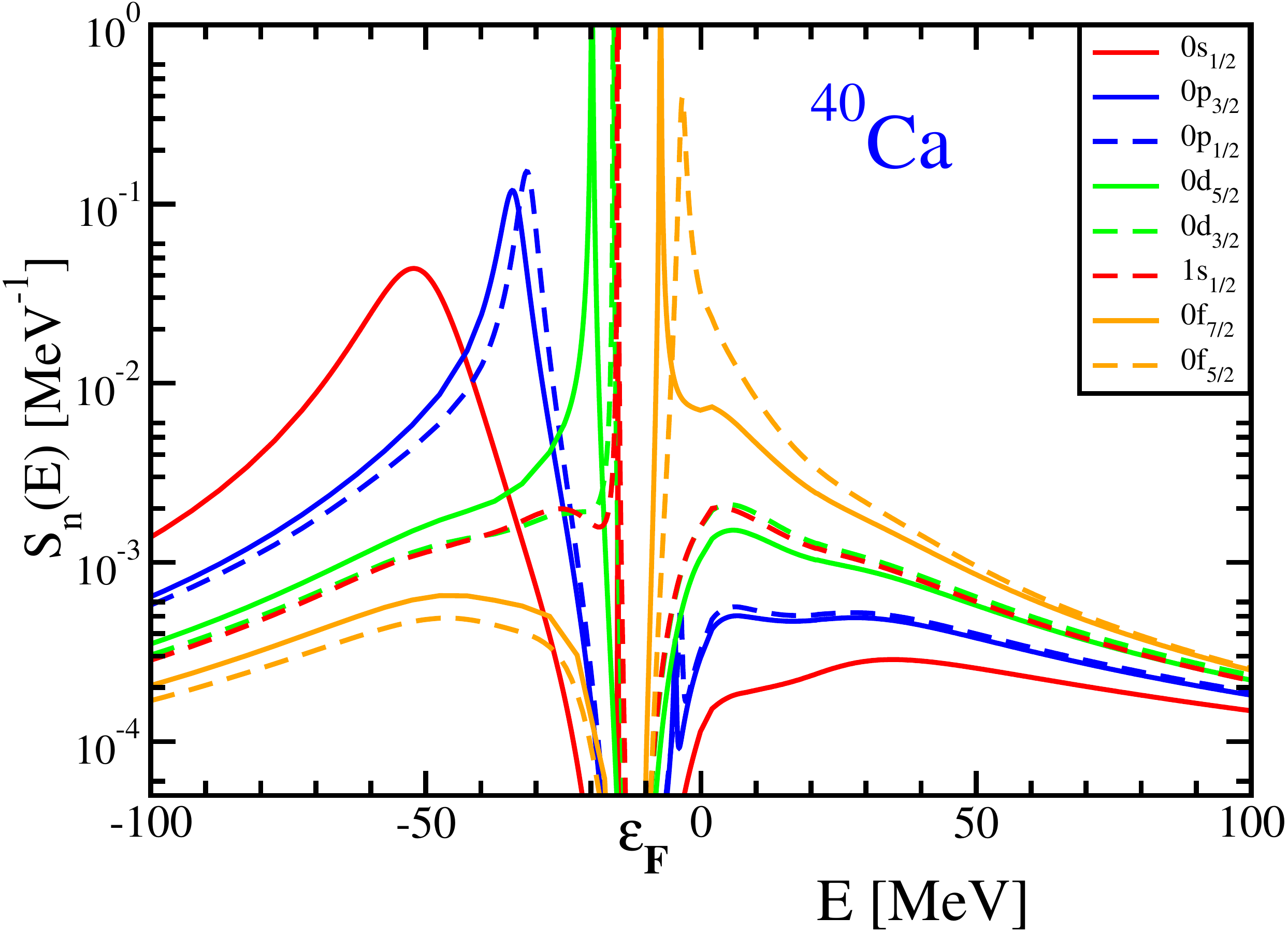}
\caption{Calculated spectral strength, both below and above the Fermi energy, for bound orbits in ${}^{40}$Ca. The spectral strength is constrained by elastic-scattering data, level structure, charge density, particle number, and the presence of high-momenta below the Fermi energy~\protect\cite{Mahzoon14}.}
\label{fig:deplE}
\end{figure}

We display in Fig.~\ref{fig:deplE} the results of the DOM spectral function for the most relevant bound orbits in ${}^{40}$Ca including the hole spectral function of Ref.~\cite{Mahzoon14}.
Because the DOM analysis assumes that the imaginary part of the self-energy starts at $\varepsilon_F$, the  spectral strength is a continuous function of the energy.
The method of solving the Dyson equation for $E<0$ is very different than that for $E>0$.
The continuity of the curves at $E=0$ confirms the numerical aspects of both of these calculations. 
Below the Fermi energy, the spectral strength contains peaks associated with the $0s_{1/2},0p_{3/2},0p_{1/2},0d_{5/2},1s_{1/2},$ and $0d_{3/2}$ orbits with narrower peaks for orbits closer to the Fermi energy.
Their strength was calculated for the overlap functions associated with the location of the peaks by solving the Dyson equation without the imaginary part but with self-consistency for the energy of the real part~\cite{Dickhoff10}.
The strength of these orbits above the Fermi energy exhibits systematic features displaying more strength when the IPM energy is closer to the continuum threshold~\cite{Michel07}.
We make this observation quantitative by listing the integrated strength according to the terms of Eq.~(\ref{eq:sumr}) in Table~\ref{Tbl:depln}. 
For the depletion we integrate from 0 to 200 MeV which corresponds to the energy domain constrained by data in the DOM.
We also include the $0f_{7/2}$ and $0f_{5/2}$ spectral functions in Fig.~\ref{fig:deplE} and corresponding results in Table~\ref{Tbl:depln} noting that the strength in the continuum from 0 to 200 MeV further rises to 0.202 and 0.320, respectively. 
From $\varepsilon_F$ to 0 the strength for these states is also included in the sum and decreases from 0.722 to 0.591, respectively.
This illustrates that there is a dramatic increase of strength into the continuum when the IPM energy approaches this threshold.
Such orbits correspond to valence states in exotic nuclei~\cite{Gade04,Jensen11,Charity14}.
The $1p_{3/2}$ and $1p_{1/2}$ spectral functions are not shown as they mimic the behavior of the $0f_{7/2}$ distribution but their presence causes the wiggles in the $0p_{3/2}$ and $0p_{1/2}$ spectral functions due to slight nonorthogonality.
%The DOM results therefore quantify the strong dependence of the amount of spectral strength in the continuum on the nearness of the bound orbit to the scattering continuum as shown in Fig.~\ref{fig:deplE} and the results in Table~\ref{Tbl:depln}.
The sensitivity to the separation from the continuum is associated with the pronounced surface absorption necessary to describe the elastic-scattering data in this energy range. 
The few percent missing strength most likely resides above 200 MeV and is associated with the effect of SRC as discussed in Ref.~\cite{Dussan11}.
\begin{table}[t]
\centering
\caption{Occupation and depletion numbers for bound orbits in ${}^{40}$Ca. The $d_{nlj}[0,200]$ depletion numbers have only been integrated from 0 to 200 MeV. The fraction of the sum rule in Eq.~(\ref{eq:sumr}) is given in the last column. }
\label{Tbl:depln}%
%\centering
\begin{tabular}{lrrc}
%  & Energy [MeV] &  \\ 
\hline
orbit  & $n_{n \ell j}$ & $d_{n \ell j}$ & $n_{n \ell j} + d_{n \ell j}$  \\
         &                      & [0,200] & $[\varepsilon_F,200]$ \\
\hline
$0s_{1/2}$ & 0.926 & 0.032  & 0.958 \\
$0p_{3/2}$& 0.914 & 0.047 & 0.961  \\
$1p_{1/2}$ &  0.906 & 0.051 &0.957  \\
$0d_{5/2}$ & 0.883 & 0.081 & 0.964  \\
$1s_{1/2}$ & 0.871 & 0.091 & 0.962  \\ 
$0d_{3/2}$ & 0.859 & 0.097 & 0.966  \\
$0f_{7/2}$ & 0.046 &  0.202 & 0.970  \\
$0f_{5/2}$ & 0.036  & 0.320 & 0.947  \\
\end{tabular}
\end{table}

\section{Results for ${}^{48}$Ca}
\label{sec-3}
Recent developments of the DOM have concentrated on the description of ${}^{48}$Ca employing nonlocal potentials~\cite{Hossein15}.
With adequate data for elastic proton scattering and the availability of an accurate charge density for this nucleus, it possible to answer the question how the properties of protons are changed when eight neutrons are added to ${}^{40}$Ca.
With recent new data for elastic neutron scattering~\cite{Charity11} and the notion that neutrons in ${}^{40}$Ca are as well described as protons due to isospin symmetry, it is also possible to assess the consequences for neutrons.
In particular, the rms radius of the neutron distribution of ${}^{48}$Ca may be receiving experimental scrutiny from a parity-violating elastic scattering experiment at Jefferson Lab~\cite{CREX13} clarifying important properties of nuclei with neutron excess relevant for the physics of neutron stars.
It may well be that available neutron data for ${}^{48}$Ca already indirectly constrain information related to the neutron properties in the ground state.
We note that the particle spectral function associated with elastic scattering shown in Fig.~\ref{fig:Sandpsi} for an $\ell = 0$ neutron in ${}^{40}$Ca clearly exhibits evidence of awareness of ground-state properties as it exhibits a two-node structure at low energy as required by the presence of mostly occupied $0s_{1/2}$ and $1s_{1/2}$ orbits.
Similar plots for $d$-waves exhibit a single node, whereas higher $\ell$-values display no node~\cite{Hossein15}.

\begin{figure}[tbp]
\centering
\includegraphics*[scale=0.45]{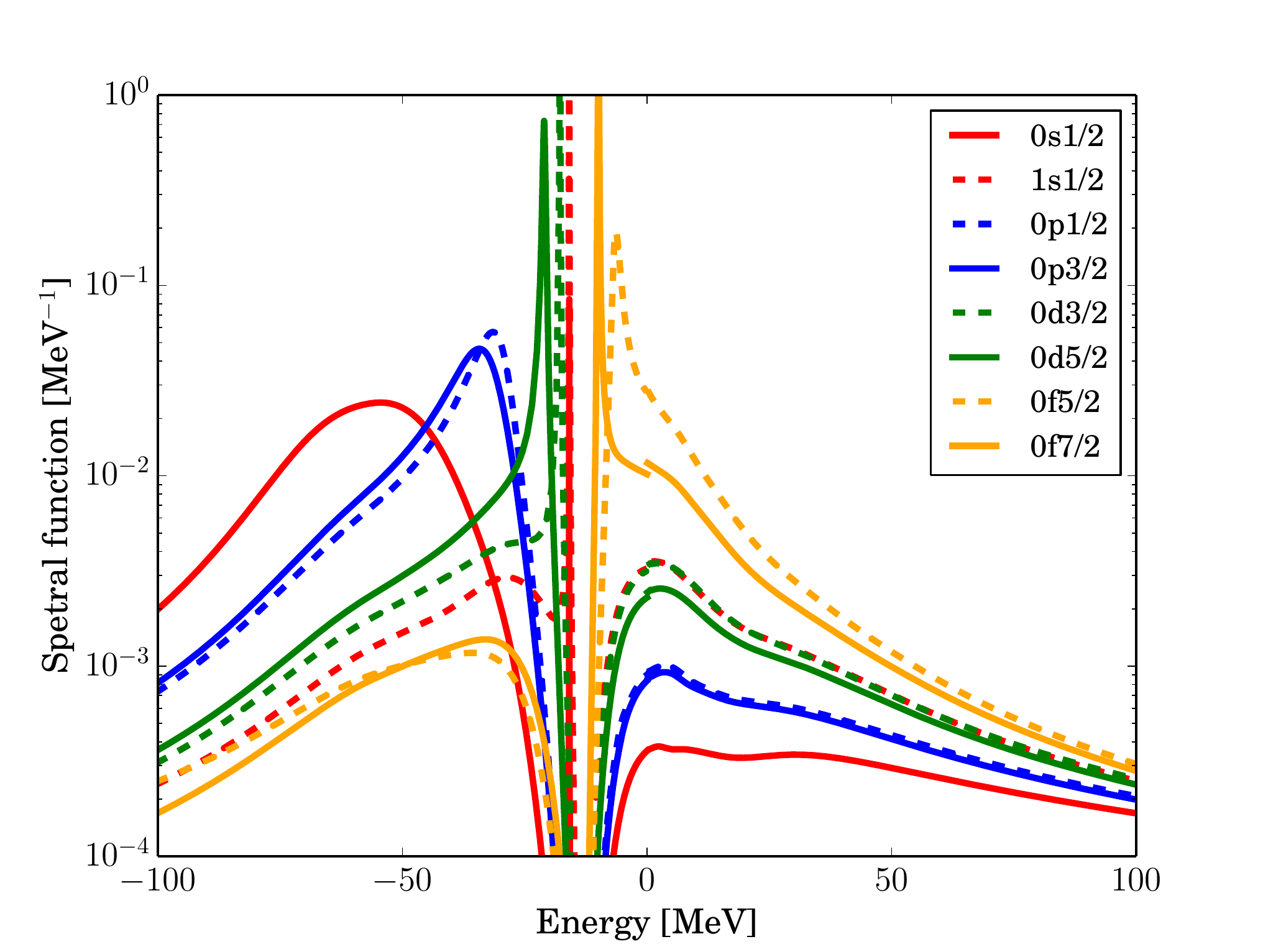}
\caption{Calculated spectral strength, both below and above the Fermi energy, for proton bound orbits in ${}^{48}$Ca. The spectral strength is constrained by elastic-scattering data, level structure, charge density, and particle number.}
\label{fig:deplEp}
\end{figure}

In complete analogy with previous local DOM analyses an asymmetry-dependent potential proportional to $(N-Z)/A$ is included for the analysis of ${}^{48}$Ca.
Nonlocality is also implemented for this component while the $N=Z$ potential associated with ${}^{40}$Ca is kept the same except for the radius parameters that are determined by ${}^{48}$Ca data~\cite{Hossein15}.
Whereas proton and neutron spectral functions for ${}^{40}$Ca as illustrated in Fig.~\ref{fig:deplE}  are equivalent except for a shift in energy to take into account the Coulomb potential for protons, this is no longer the case for
${}^{48}$Ca.
It is therefore necessary to treat proton elastic scattering in momentum space in such a way that the relevant information can be obtained but the full treatment of the Coulomb potential is avoided.
The appropriate method has been explored in Refs.~\cite{Deltuva05a,Deltuva05b,Deltuva08} and we have adopted it here by screening the Coulomb potential in such a way that its contribution is completely included in the domain where the nuclear potential acts~\cite{Hossein15}.
The results for the proton spectral functions for ${}{48}$Ca are shown in Fig.~\ref{fig:deplEp} and exhibit a slight discontinuity at $E=0$ where the treatment of the Coulomb potential at positive energy is particularly delicate.
We note that the $0s_{1/2}$ spectral function in Fig.~\ref{fig:deplEp} has a wider peak than the corresponding result for ${}^{40}$Ca as the width of this orbit is not properly constrained in ${}^{48}$Ca. One would not expect a substantial difference between these nuclei however for this quantity.
Similarly, there are no experimental high-momentum removal constraints for this nucleus which would be very helpful in clarifying the change from $N=20$ to $N=28$ for protons with high momenta. 

\begin{table}[b]
\centering
\caption{Spectroscopic factors for orbits in the immediate vicinity of the Fermi surface in ${}^{40}$Ca and ${}^{48}$Ca. Proton and neutron spectroscopic factors in ${}^{40}$Ca are identical. }
\label{Tbl:sfac}%
%\centering
\begin{tabular}{lrrc}
%  & Energy [MeV] &  \\ 
\hline
orbit  & ${}^{40}$Ca & $p\ {}^{48}$Ca & $n\ {}^{48}$Ca  \\
\hline
$1s_{1/2}$ & 0.78 & 0.71 & 0.83  \\ 
$0d_{3/2}$ & 0.76 & 0.65 & 0.80  \\
$0f_{7/2}$ & 0.73 &  0.59 & 0.84   \\
\end{tabular}
\end{table}
Near the Fermi energy sufficient experimental data exist and it is possible to make a quantitative comparison of spectroscopic factors between these two nuclei as shown in Table~\ref{Tbl:sfac}.
We first note that the spectroscopic factors of the $0d_{3/2}$ and $1s_{1/2}$ orbits in ${}^{40}$Ca are larger
by about 10-15\% than the numbers extracted in the analysis of the $(e,e'p)$ reaction of Ref.~\cite{Kramer89}.
The relativistic analysis of the $(e,e'p)$ reaction in Ref.~\cite{Udias95} clarified that the treatment of nonlocality  leads to different distorted proton waves as compared to conventional non-relativistic optical potentials, yielding about 10-15\% larger spectroscopic factors.
No matter where the absolute values are situated, it is clear from Table~\ref{Tbl:sfac} that proton spectroscopic factors for these orbits are smaller in ${}^{48}$Ca whereas neutron ones are larger.
Since these results are strongly constrained by experimental elastic scattering data, it appears to provide some support for the notion that the spectroscopic factors of the more strongly bound valence orbits of the minority nucleon decrease whereas the ones for the less bound majority nucleons increases with increasing nucleon asymmetry~\cite{Gade04}.
For the particle spectroscopic factor of the proton $0f_{7/2}$ orbit a similar observation can be made.
We note that for neutrons this orbit switches from being mostly empty to mostly occupied between these two nuclei but also in this case the spectroscopic factor increases from ${}^{40}$Ca to ${}^{48}$Ca.
For protons it is quite clear what the explanation of this decrease of the valence spectroscopic factors amounts to.
Elastic scattering reaction cross section data clearly show more surface absorption for protons in ${}^{48}$Ca than for ${}^{48}$Ca~\cite{Charity07}.
The correspondingly larger imaginary part of the potential will therefore remove additional strength in ${}^{48}$Ca from valence proton orbits as compared to ${}^{40}$Ca leading to an unavoidable reduction of the spectroscopic factors.

\begin{figure}
\centering
\includegraphics[width=9cm,clip]{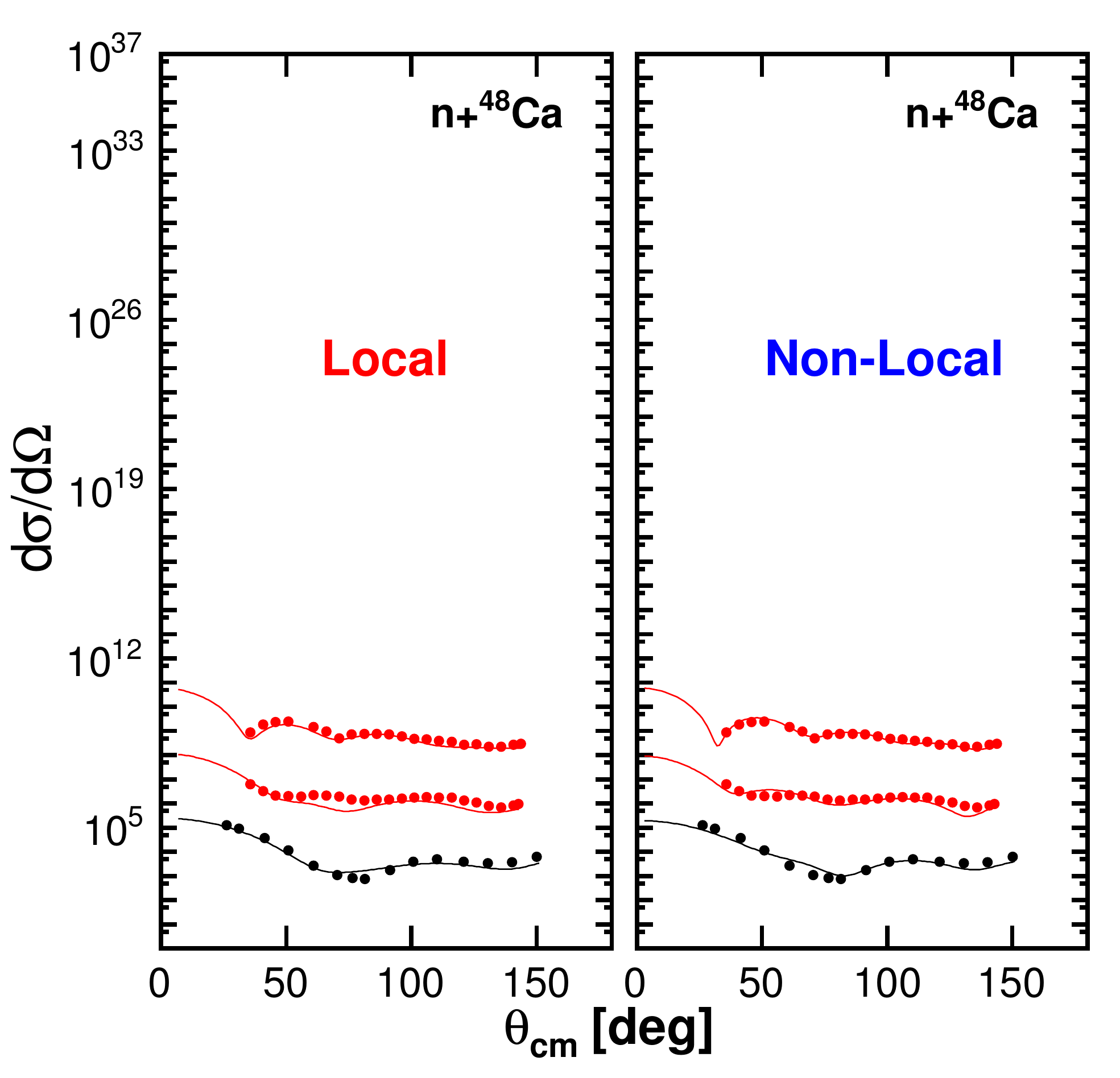}
\caption{Calculated and experimental elastic-scattering angular distributions of the differential cross section for $n+^{48}$Ca in the local (left panel)~\cite{Charity11} and nonlocal DOM implementation~\cite{Hossein15}. Data for each energy are offset for clarity. References to the data can be found in Ref.~\cite{Charity11}.}
\label{fig-7}       % Give a unique label
\end{figure}

In a previous local DOM analysis of elastic neutron scattering from ${}^{48}$Ca~\cite{Charity11} some difficulties were encountered in describing the three sets of available differential cross section.
As shown in Fig.~\ref{fig-7} the nonlocal fit is apparently capable of describing these data quite satisfactorily when the fits are compared side by side.
The fit of the proton elastic scattering data is of the same quality as achieved with the local DOM~\cite{Hossein15}.
Results for the fits to reaction cross sections for both protons and neutrons and total cross sections for neutrons are also well represented in the nonlocal DOM~\cite{Hossein15}. 

\begin{figure}
\centering
\includegraphics[width=9cm]{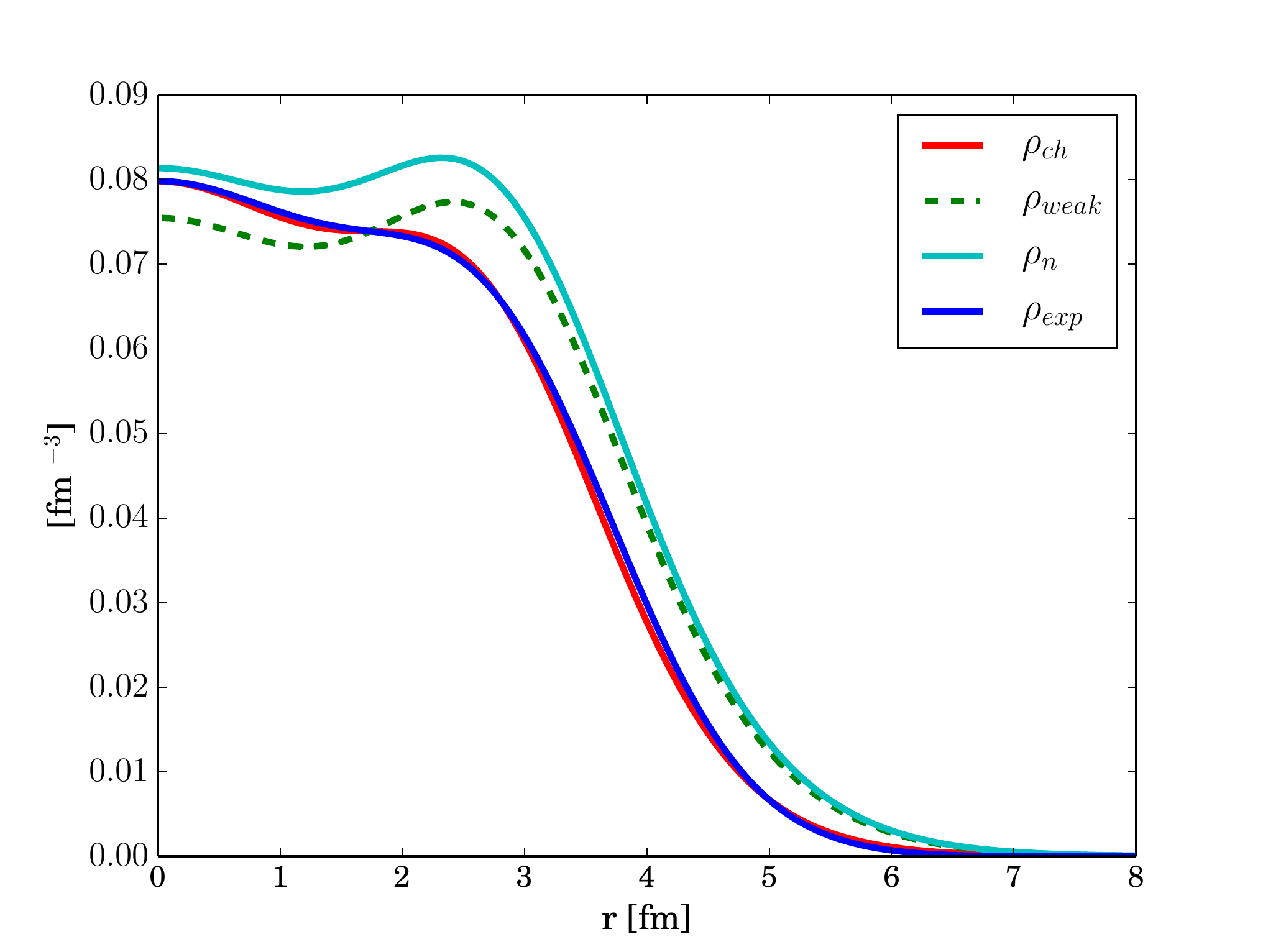}
\caption{Comparison of calculated and experimental charge density for ${}^{48}$Ca.
The calculated weak charge distribution (dashed line as well as the neutron matter distribution are plotted as well~\cite{Hossein15}.}
\label{fig-8}       % Give a unique label
\end{figure}

Finally, results for the charge density of ${}^{48}$Ca are compared with the experimental results~\cite{deVries1987} together with results for the neutron distribution and the weak charge in Fig.~\ref{fig-8}.
The quality of the charge density fit is similar as for ${}^{40}$Ca shown in Fig.~\ref{fig-3}.
As the main difference between the proton and neutron distribution is associated with $0f_{7/2}$ neutrons, it is reasonable to expect the interior densities to be comparable as shown in Fig.~\ref{fig-8}.
This observation suggests that at higher energy where elastic scattering for protons is dominated by volume physics the protons interact with a similar composition of nucleons in ${}^{40}$Ca and ${}^{48}$Ca.
This notion is confirmed by comparing volume integrals of the imaginary part of the potentials as a function of energy for these nuclei~\cite{Hossein15}.

The result for the neutron skin accordingly is quite large $R_n-R_p = 0.249 \pm 0.023$ fm considerably larger than a prediction of an \textit{ab initio} coupled-cluster calculation~\cite{Hagen2015} and larger than most mean-field results compiled in Ref.~\cite{skingraph}.
The error quoted for the nonlocal DOM result is obtained by scrambling the available elastic scattering data with the assigned experimental errors for neutron scattering on ${}^{48}$Ca.
A clear impetus therefore exists to extend our nonlocal DOM analysis to a description of ${}^{208}$Pb where a measurement for the neutron skin has already been reported~\cite{prex} but unfortunately with a large error bar.
%\newpage
%For one-column wide figures use syntax of figure~\ref{fig-1}

%For two-column wide figures use syntax of figure~\ref{fig-2}
%\begin{figure*}
%\centering
%\vspace*{5cm}       % Give the correct figure height in cm
%\label{fig-4}       % Give a unique label
%\end{figure*}
%
%For figure with sidecaption legend use syntax of figure
%\begin{figure}
%\centering
%\sidecaption
%\includegraphics[width=5cm,clip]{MissingNEW.eps}
%\label{fig-5}       % Give a unique label
%\end{figure}
%
%For tables use syntax in table~\ref{tab-1}.
%\begin{table}
%\centering
%\label{tab-1}       % Give a unique label
% For LaTeX tables you can use
%\begin{tabular}{lll}
%\hline
%first & second & third  \\\hline
%number & number & number \\\hline
%\end{tabular}
% Or use
%\vspace*{5cm}  % with the correct table height
%\end{table}

\section{Conclusions and outlook}
\label{sec-4}
The overview presented in this contribution clearly demonstrates that it is possible to fruitfully employ the link between nuclear reactions and structure as provided by the DOM.
The biggest step forward in recent years has been the implementation of fully nonlocal and dispersive optical potentials for Ca isotopes allowing for an accurate fit of ground-state properties like the nuclear charge density. The analysis of scattering data employing the particle spectral density furthermore documents the depletion of orbits that are mostly occupied in the ground state.
Apparently the present DOM implementation is also capable of \textit{predicting} ground-state properties like the neutron skin with suggestive results that clamor for experimental probing. 
Future work on ${}^{208}$Pb is clearly called for.
In addition it is important to point out that the nonlocal analysis leads to distorted waves that are quite different from those obtained from local potentials potentially leading the new results when \textit{e.g.} transfer reactions are analyzed with nonlocal potentials~\cite{Ross2015}.
DOM ingredients based on local potentials already generate more consistent results for the analysis of transfer reactions~\cite{Nguyen11}.
\section*{Acknowledgement}
The material presented here is based upon work supported by the U.S. Department of Energy, Office of Science, Office of Nuclear Physics under Award number DE-FG02-87ER-40316 and by the U.S. National Science Foundation under grant PHY-1304242.
At this point I would like to acknowledge the work of Hossein Mahzoon whose thesis research has been instrumental in generating most of the results reported in this contribution. The work of other collaborators is also very much appreciated.
%
% BibTeX or Biber users please use (the style is already called in the class, ensure that the "woc.bst" style is in your local directory)
\bibliography{MBT18}
%\begin{thebibliography}{}
%Journal Author, Journal \textbf{Volume}, page numbers (year)
%Book Author, \textit{Book title} (Publisher, place, year) page numbers
% etc
%\end{thebibliography}

\end{document}